\documentstyle[12pt]{article}
\begin{document}
\begin{center}
{\bf COLLECTIVE DYNAMICS OF A DOMAIN WALL --- AN OUTLINE } \\
$ \;$ \\
by \\
$ \;$ \\
{\bf H. Arod\'{z}} \\

Institute of Physics, Jagellonian University, \\
Reymonta 4,  30-059 Cracow, Poland  
\end{center}
\vspace*{2cm}

We shall consider domain walls in a relativistic field-theoretical
model defined by the following Lagrangian
\begin{equation}
{\cal L} = -\frac{1}{2}\eta_{\mu\nu}\partial^{\mu}\Phi \partial^{\nu}\Phi
-\frac{\lambda}{2}(\Phi^{2}-\frac{M^2}{4\lambda})^2, \label{1}
\end{equation}
where $\Phi$ is a single real scalar field, $(\eta_{\mu\nu})$=diag(-1,1,1,1)
is the metric in Minkowski space-time, and $\lambda,M$ are positive
parameters. 

Euler-Lagrange equation corresponding to Lagrangian (1) has the 
particular time-independent solution
\begin{equation}
\Phi(x^{\mu}) = \Phi_0 \tanh(\frac{x^3}{2l_0})
\end{equation}
which describes a static, planar domain wall stretched along the $x^3=0$
plane. Here $\Phi_0 = M/(2\sqrt{\lambda}) $ denotes one of the two vacuum
values of the field $\Phi$ --- the other one is equal to $ - \Phi_0$. The
parameter $M$ can be identified with the mass of the scalar particle related
to the field $\Phi$. The corresponding Compton length $l_0 = M^{-1}$
gives the physical length scale in the model.
Energy density for the planar domain wall (2) is
exponentially localised in a vicinity of the $x^3 =0$ plane. The transverse
width of the domain wall is of the order $2l_0$.

The issue is time evolution of a non-planar domain wall. Such domain walls
can be infinite, consider, e.g., locally deformed  planar domain wall
or a cylindrical domain wall. They can also be finite closed, e.g., like
a sphere or a torus. We shall restrict our considerations to a single
large, smooth domain wall. Such a domain wall is defined by a set of
conditions which provide  Lorentz and reparametrisation invariant
formulation of the heuristic requirement
that  $l_0^2 R_{1,2} \ll 1$ at each point of the domain wall, where 
$R_1, R_2$ denote local main curvature radia in a
local rest frame of the considered infinitesimal piece of the domain
wall. In this case it is possible to develop the presented below 
perturbative approach to the dynamics of the domain wall.
In other cases, e.g. when  $l_0/R_i \sim 1$ at certain
points of the domain wall, the only practical tool is numerical analysis.

Physical idea underlying the expansion in the width is rather simple.
The domain wall is a stable solitonic object. One expects that  each
piece of the large, smooth domain wall in the local co-moving reference frame
does not differ much from the planar domain wall. The differences, which are
due to curvature of the domain wall, can be calculated perturbatively.

The tricky point is to find a convenient reference frame co-moving with
the domain wall. From the seminal papers \cite{1,2}, which were devoted
to dynamics of vortices but this does not make an essential difference here,
we know that the co-moving reference frame should be based on a relativistic
membrane. In the leading approximation \cite{1,2} the membrane is of
Nambu-Goto type, and at each instant of time it coincides with the surface
on which the scalar field $\Phi$ vanishes. In the following we will call
this surface {\it the core} of the domain wall. For instance, in the
case of the static domain wall (2) the core is given by the plane $x^3=0$.
When calculating corrections to the leading approximation one can still
adhere to this identification of the membrane with the core,
see, e.g., papers \cite{3,4,5}. The price for
this is that the membrane is no longer of the Nambu-Goto type. Effective
action for such membrane contains higher derivatives with respect to time,
and probably is nonlocal \cite{6}.
The reason for these unpleasant features
is that the points at which the scalar field vanishes  do not constitute a
physical object (the physical object is the domain wall itself and not the
core), and therefore the core, being a purely mathematical construct, can
have strange from the physical point of view equation of motion.
In papers \cite{7,8} we have proposed to use a membrane which is
of Nambu-Goto type to all orders. This membrane coincides with the
core only at the initial instant of time.

Our description of the dynamics of the domain wall \cite{7,8} involves
the Nambu-Goto membrane and certain additional 2+1-dimensional fields
defined on the membrane. They obey nonlinear equations of motion
which are second order partial differential equations, and they describe
time evolution of the domain wall completely. Transverse profile of the
domain wall, that is dependence of the scalar field $\Phi$ on a variable
changing in the direction perpendicular to the domain wall, is uniquely and
explicitly given once the evolution of the Nambu-Goto membrane and of
the 2+1-dimensional fields is found.
The dependence on the transverse variable is calculated perturbatively, and
one should observe that it is the case of a singular perturbation. Our
approach to constructing the perturbative expansion follows a method used in
condensed matter physics \cite{9}. Because of the explicit
dependence on the transverse variable the
dynamics of the domain wall has been reduced to dynamics of the collective
degrees of freedom given by the Nambu-Goto membrane and the fields on it.
In the following we shall present basic steps of the approach developed in
papers \cite{7,8}.
\\
{\bf Step 1. The co-moving coordinate system.} \\
We introduce a surface $S$
co-moving with the domain wall, that is {\it the co-moving membrane}.
It does not have to coincide with the core,
except at the initial instant of time. The world-volume of $S$, denoted
by $\Sigma$, is parametrised as follows
\begin{equation}
\Sigma \ni  ( Y^{\mu} ) (u^a) = ( \tau, Y^i( u^a) ).
\end{equation}
We use the notation $(u^a)_{ a=0, 1, 2} = (\tau, \sigma^1, \sigma^2)$,
where $\tau$ coincides with the laboratory frame time $x^{0}$, while
$\sigma^1, \sigma^2$ parametrise   $S$ at each
instant of time. The index $i=1,2,3$ refers to the spatial components
of the four-vector. The points of the co-moving membrane $S$ at the
instant $\tau_0$ are given by $ (Y^i)(\tau_0, \sigma^1, \sigma^2) $.
The coordinate system $(\tau,\sigma^1,\sigma^2,\xi)$ co-moving with
the domain wall is defined by the formula
\begin{equation}
x^{\mu} = Y^{\mu}( u^a) + \xi \: n^{\mu}( u^a),
\end{equation}
where $x^{\mu}$ are Cartesian (laboratory frame) coordinates in Minkowski
space-time, and $(n^{\mu})$ is a normalised space-like four-vector
orthogonal to  $\Sigma$ in the covariant sense,
\[
n_{\mu}( u^a )Y^{\mu}_{,a}( u^a ) = 0,
  \;\;\;\;\; n_{\mu}n^{\mu}=1,
\]
where $Y^{\mu}_{,a}\equiv \partial Y^{\mu}/ \partial u^a$. The three
four-vectors $Y_{,a}$ are tangent to $\Sigma$. $\xi$ is the transverse
variable. Definition
(4) implies that $\xi$ and $u^a$ are Lorentz scalars. In the co-moving
coordinates the co-moving membrane is described by the
condition $\xi=0$.  For points lying on  $S$
the parameter $\tau$ coincides with the laboratory time $x^{0}$, but for
$\xi \neq 0$  in general $\tau$ is not equal to  $x^0$.

The extrinsic curvature coefficients $K_{ab}$ and induced
metrics $g_{ab}$ on $\Sigma$ are defined by the following formulas:
\[
K_{ab} = n_{\mu} Y^{\mu}_{,ab}, \;\;\;\;
  g_{ab} = Y^{\mu}_{,a} Y_{\mu,b},
\]
where $a,b=0,1,2$. The covariant metric tensor in the new coordinates
has the following form 
\[
[G_{\alpha\beta}] = \left[ \begin{array}{lr} G_{ab} & 0 \\
0 & 1 \end{array} \right], 
\]
where $\alpha,\beta=0,1,2,3$;\  $\;\;\;\alpha=3$ corresponds to the $\xi$
coordinate; and
\[
G_{ab}=N_{ac} g^{cd} N_{db},\;\;\; 
N_{ac} = g_{ac} - \xi K_{ac}.
\]
Thus, $G_{\xi\xi}=1, \; G_{\xi a}=0$.
Straightforward computation gives  
\[
\sqrt{-G} = \sqrt{-g} \; h(\xi, u^a),
\]
where as usual  $g = det[g_{ab}]$, $\;\;$
$G = det[G_{\alpha\beta} ]$, and
\[
h(\xi, u^a) =
1 - \xi K^{a}_{a} +\frac{1}{2} \xi^2 (K^{a}_{a} K^{b}_{b} - K^{b}_{a}
K^{a}_{b}) -\frac{1}{3} \xi^3  K^{a}_{b} K^{b}_{c} K^{c}_{a}.
\]
For raising and lowering the latin indices of the extrinsic 
curvature coefficients we use the induced metric tensors 
$g^{ab},\;\; g_{ab}$. 

The inverse metric tensor $G^{\alpha \beta}$ is given by the formula
\[
[G^{\alpha \beta}] = \left[ \begin{array}{lr} G^{ab} &  0 \\
0 &   1 \end{array} \right],
\]
where
\[
G^{ab}= (N^{-1})^{ac} g_{cd} (N^{-1})^{db}.
\]
$ N^{-1}$ is just the matrix inverse to $[N_{ab}]$. It has the upper indices
by definition.

In general, the coordinates $( u^a, \xi)$ are defined locally,  in a 
vicinity of the world-volume $\Sigma$ of the membrane.
Roughly speaking, the allowed range of the $\xi$ coordinate is determined
by the smaller of the two main curvature radia of the membrane
in a local rest frame. We assume that this curvature radius is
sufficiently large so that on the
outside of the region of validity of the co-moving coordinates there are 
only exponential tails of the domain wall, that is that the field $\phi$ is
exponentially close to one of the two vacuum solutions.
\\
{\bf Step 2. Field equation in the co-moving coordinates.} \\
It is convenient to rescale the field $\Phi$ and the coordinate $\xi$,
\[
\Phi(x^{\mu}) =  \frac{M}{2\sqrt{\lambda}} \phi(s, u^a),
\;\; \xi = \frac{2}{M} s,
\]
where $\phi$ and $s$ are dimensionless. We also extract from the
scalar field its component living on the co-moving
membrane and treat it separately from the remaining part of the
scalar field.  To this end we write the identity
\begin{equation}
\phi(s, u^a) = B(u^a) \psi(s) + \chi(s, u^a),
\end{equation}
where 
\begin{equation}
 B(u^a) \stackrel{df}{=} \phi(0, u^a)
\end{equation}
is the component of the scalar field living on the co-moving membrane, and
\begin{equation}
\chi \stackrel{df}{=} \phi(s, u^a) - B(u^a) \psi_0(s)
\end{equation}
is the remaining part.
The auxiliary, {\it fixed} function $\psi_0(s)$ depends on the variable
$s$ only. It is smooth, concentrated around $s = 0$, and
\begin{equation}
\psi_0(0) = 1.
\end{equation}
It follows that 
\begin{equation}
\chi(0, u^a) = 0.
\end{equation}
The best choice for $\psi_0(s)$ is given by formula \cite{8}
\[ \psi_0(s) = \frac{1}{\cosh^2(s)}.  \]
Next we  derive Euler-Lagrange equations
by taking independent variations of $B(u^a)$ and $\chi$. The
variation $\delta \chi$  has to respect condition (9), hence
\[
\delta \chi(0, u^a) = 0.
\]
Because of this condition,  variation of the action functional
\[ {\cal S} = \frac{2}{M} \int ds d^3u \; \sqrt{-g} h(s, u^a) {\cal L}
\]
with respect to $\chi$ gives Euler-Lagrange equation in the regions
$s < 0$ and $s > 0$. It has the following form
\begin{eqnarray}
\lefteqn{\frac{2}{M^2}  \frac{1}{\sqrt{-g}  }
\partial_a [\sqrt{-g} h G^{ab} \partial_b ( B \psi_0 +
 \chi ) ]   } \\
& & + \frac{1}{2 }  \partial_s [ h \partial_s ( B \psi_0 + \chi)]
 + h (B \psi_0 + \chi) [1 - (B \psi_0 + \chi)^2] = 0. \nonumber
\end{eqnarray}
At $s = 0$ there is no Euler-Lagrange equation corresponding to the
variation $\delta\chi$. Instead, we have the condition (9). Equation (10)
should be solved in the both regions separately, with (9) regarded as
a part of boundary conditions for $\chi$. To complete the boundary
conditions we also specify the behaviour of $\chi$ for $|\xi|$ much larger
than the characteristic length $l_0$, that is for $|s| \gg 1$:
we shall seek a solution such that
$\chi$  is exponentially close to +1 for $s \gg 1$ , while for $s \ll -1$
it is exponentially close to $-1$.

At this stage of considerations Eq.(10) should not be extrapolated
to $s = 0$. For example, the l.h.s. of it could  have a $\delta(s)$-type
singularity. It would occur if $\chi$ was smooth for $s>0$ and for $s<0$
but had a spike at $s = 0$.

In addition to Eq.(10) we also have the Euler-Lagrange equation
corresponding to variations of $B(u^a)$. This equation has the
following form
\begin{eqnarray}
\lefteqn{\frac{2}{M^2} \int ds \; \frac{1}{\sqrt{-g} } 
\partial_a \left[\sqrt{-g} h G^{ab} \partial_b ( B \psi_0 +
\chi ) \right]  \psi_0 }  \\
&  & - \frac{1}{2} \int ds \; h \partial_s \psi_0 \partial_s
( B \psi_0 + \chi )
 \nonumber \\
& &  + \int ds \; h \psi_0 ( B \psi_0 +  \chi ) \left[ 1 - ( B \psi_0
 + \chi )^2
\right] = 0. \nonumber
\end{eqnarray}
Here and in the following we use $\int ds$ as a shorthand for the definite
integral $\int_{-\infty}^{+\infty} ds$.

Using equations (10) {\it  and} (11) one can prove
that $\chi$ does not have the spike at $s = 0$.
It follows that Eq.(10) is obeyed by $\chi$ also at $s = 0$, and that now
Eq.(11) can be obtained by multiplying Eq.(10) by $\psi_0(s)$ and integrating
over $s$. Therefore we may concentrate on solving Eq.(10).
\\
{\bf Step 3.  Expansion in the width.} \\
First we solve Eq.(10) in the leading approximation
obtained by putting $1/M = 0$. The equation is then reduced to
\begin{equation}
\frac{1}{2}  \partial^2_s \phi^{(0)} +  \phi^{(0)} [1- (\phi^{(0)})^2] = 0,
\end{equation}
where
\[ \phi ^{(0)} = B^{(0)} \psi_0 + \chi ^{(0)}.  \]
Equation (12) does not contain derivatives with respect to time,
in spite of the fact that it  is  supposed to approximate the evolution
equation (10).  This annoying fact is due to the singular character of
the perturbation given in Eq.(10) by the terms proportional to positive
powers of  $1/M$.
We shall see that time evolution is obtained indirectly, from
consistency conditions.

Equation (12)  has the following particular, well-known solution
\begin{equation}
\phi^{(0)} = \tanh s.
\end{equation}
This solution together with conditions
(8), (9) gives
\begin{equation}
B^{(0)} =0, \;\;\; \chi^{(0)} = \tanh s.
\end{equation}
Notice that $\phi^{(0)}$ has the same form as the
planar domain wall (2) --- in this way we realise the idea that in the
co-moving reference frame the domain wall does not differ much from the
planar domain wall.

The solution (13) in the co-moving coordinates does not determine
the field $\phi$ in the laboratory frame because we do not know yet the
position of the co-moving membrane with respect to the laboratory frame.
Equations (10), (11)  yield an equation
for the co-moving membrane, otherwise they would not form the complete
set of evolution equations for the field $\Phi$.
In fact, we shall see that the first order terms in Eq.(10)
imply  Nambu-Goto equation for the membrane.

The expansion in the width has the form
\begin{equation}
\chi(s, u^a) = \tanh s + \frac{1}{M} \chi^{(1)}(s,u^a) +
\frac{1}{M^2} \chi^{(2)}(s,u^a)   + \frac{1}{M^3}  \chi^{(3)}(s,u^a) + ... ,
\end{equation}
\begin{equation}
B(u^a) =  \frac{1}{M} B^{(1)}(u^a) +  \frac{1}{M^2} B^{(2)}(u^a)
  + \frac{1}{M^3}  B^{(3)}(u^a) + ... ,
\end{equation}
where we have taken into account the zeroth order results (14).
The expansion parameter is $1/M$
and not $1/M^2$ because $1/M$ in the first power appears in $h$ and
$G^{ab}$  functions after passing to the $s$ variable.
In order to obey the condition (9), and to ensure the proper
asymptotics of $\chi$ at large $|s|$ we assume that for $n \geq 1$
\begin{equation}
\chi^{(n)}(0, u^a) = 0, \;\;\; \lim_{s \rightarrow \pm \infty}
\chi^{(n)} = 0.
\end{equation}
Inserting the perturbative Ansatz (15, 16) in Eqs.(10, 11),
expanding the l.h.s.'s of them in powers of $1/M$, and equating to zero
coefficients in front of the powers of 1/M we obtain a sequence of 
linear, inhomogeneous equations for $\chi^{(n)}(s, u^a), B^{(n)}(u^a)$ with
$n\geq 1$.

In particular, Eq.(10) expanded in the powers of $1/M$ gives
equations of the type
\begin{equation}
\hat{L} \chi^{(n)} = f^{(n)},
\end{equation}
where the source term $ f^{(n)} $ is determined by the lower order terms
in $\chi$ and $B$, and
\[
\hat{L} \stackrel{df}{=} \frac{1}{2} \partial_s^2 + 1 - 3 (\chi^{(0)})^2.
\]
Explicit solution of Eq.(18) is given by the formula
\begin{equation}
\chi^{(n)}(s) = 2 \psi_1(s) \int_{-\infty}^s dx \; \psi_0(x) f^{(n)}(x)
- 2 \psi_0(s) \int_0^s dx \; \psi_1(x) f^{(n)}(x).
\end{equation}
This solution obeys the boundary conditions (17).

Obviously, we assume that all  proportional to positive powers of $1/M$ 
terms in Eqs.(10, 11)  are small. For this it is not sufficient that
the extrinsic curvatures are small, that is that  $l_0 K^a_b \ll 1$.
We have also to assume that the
derivatives $\chi^{(n)}_{,a}, B^{(n)}_{,a}$ are not proportional to M.
It is not the case, for example, if $\chi$ and $B$ contain modes 
oscillating with a frequency $\sim M$. They would give positive powers of
$M$ upon differentiation with respect to $u^a$. If such oscillating
components were present the counting of powers of $1/M$ would no longer
be so straightforward as we have assumed. This assumption excludes radiation
modes as well as massive excitations of
the domain wall. Therefore, the approximate solution we
obtain gives what we may call {\it the basic curved domain wall}.
To obtain more general domain wall solutions one would have to
generalize appropriately the approximation scheme. Actually, the fact
that such particular radiationless, unexcited curved domain wall exists
is a prediction coming from the $1/M$ expansion. The expansion
yields domain walls of concrete transverse profile --- the dependence
on $s$ is explicit in the approximate solution we construct even at
the initial instant of time.  We may choose the initial position and
velocity of points of the membrane but the dependence of the scalar
field on the variable $s$ at the initial time is given by formulas (19). 
This unique profile is characteristic for the basic curved domain wall.
\\
{\bf Step 4. The consistency conditions.} \\
The first order terms in Eq.(10) give the following equation
\begin{equation}
 \hat{L} \chi^{(1)} =   K_a^a  \partial_s \chi^{(0)},
\end{equation}
where $\chi^{(0)}$ is given by the second of formulas (14).

The most important point in our approach
is the observation that operator $\hat{L}$ has a zero-mode, that is
the normalizable solution 
\[
\psi_0(s) = \frac{1}{\cosh^2s} 
\]
of the homogeneous equation
\begin{equation}
\hat{L}\psi_0 = 0.
\end{equation}
Notice that $\psi_0 = \partial_s \chi^{(0)}$ --- this means that the
zero-mode $\psi_0$ is related to the translational invariance of Eq.(12)
under $s \rightarrow s$ + const.
The presence of the zero-mode implies the consistency conditions. For
example, let us multiply Eq.(20) by $\psi_0$ and integrate over $s$.
It is easy to see that $\int \psi_0 \hat{L} \chi^{(1)}$
vanishes because of (21), and we obtain the following condition
\[
K^a_a \int ds \; \psi_0(s) \partial_s \chi^{(0)}(s) = 0
\]
which is equivalent to
\begin{equation}
K^a_a = 0.
\end{equation}
Eq.(22) coincides with the well-known Nambu-Goto equation. It determines
the motion of the co-moving membrane, that is the functions $Y^i(u^a)$,
i=1,2,3, once initial data are fixed. When we know these functions we
can calculate the extrinsic curvature coefficients $K_{ab}$ and the
metric $g_{ab}$.  Review of properties of relativistic Nambu-Goto
membranes can be found in, e.g., \cite{9}.

Due to Nambu-Goto equation (22) the r.h.s. of Eq.(20)
vanishes and the resulting homogeneous equation
\[
\hat{L} \chi^{(1)} = 0
\]
with the boundary conditions (17) has only the trivial solution
\[
\chi^{(1)} = 0.
\]
Notice that vanishing $\chi^{(1)}$ does not mean that
the first order correction to the total field $\phi$ also vanishes ---
there is the first order contribution equal to $B^{(1)} \psi_0/M$.
It does not vanish on the co-moving membrane that is at $s = 0.$

Analogous reasoning can give nontrivial consistency conditions also for
equations (18) with $n>1$. For some $n$, e.g. $n=2$ we obtain only the
trivial identity 0 = 0.

Equation (11)  expanded in powers of $1/M$ gives equations for $B^{(n)}$
coinciding with the consistency conditions. This follows from the fact that
both Eq.(11) and the consistency conditions are obtained by multiplying
Eq.(10) by the zero-mode $\psi_0$ and
next integrating over $s$. Euler-Lagrange equation (11) can be regarded
as  generating equation for the consistency conditions.

Equations (10), (11) in the first order do not give any restriction
on the function $B^{(1)}$. Equation for $B^{(1)}$ follows from the
third order terms in Eq.(11):
\begin{equation}
\frac{1}{\sqrt{-g} } 
\partial_a( \sqrt{-g} g^{ab}\partial_bB^{(1)}) + 
(\frac{\pi^2}{4} - 1)  K^a_b K^b_a  B^{(1)}
+ \frac{9}{35} (B^{(1)})^3  =
(\frac{\pi^2}{6} - 1)  K^a_b K^b_c K^c_a,
\end{equation}
This situation is typical for singular perturbation theories of which the 
$1/M$ expansion is an example --- higher order equations imply restrictions 
(the consistency conditions) for the lower order contributions \cite{10}.
\\
{\bf Step 5. Initial data.} \\
At this point we have the complete set of equations determining the
evolution of the domain wall in the $1/M$ expansion. 
Each of Eqs.(10), (11), (22) describes
different aspect of the dynamics of the curved domain wall.
Expanded in the positive powers of $1/M$ Eq.(10) determines dependence
of $\chi$ on $s$. Because the term $B\psi_0$ in formula (5) has explicit
dependence on $s$, we may say that Eq.(10) for $\chi$ fixes the
transverse profile of the domain wall.

Equation (11) determines the  $B^{(n)}(u^a)$ functions,
which can be regarded
as (2+1)-dimensional scalar fields defined on $\Sigma$ and having
nontrivial nonlinear dynamics. The extrinsic curvature $K_{ab}$ of
$\Sigma$ acts as 
an external source for these fields.  The  fields $B^{(n)}$ can propagate
along $\Sigma$. One may regard this effect as {\it causal} propagation
of deformations which are introduced by the extrinsic curvature.

Finally, Nambu-Goto equation (22) for the co-moving membrane determines
the evolution of the shape of the domain wall.

Equation (22) for the co-moving membrane and equations for $B^{(n)}$ obtained
from the consistency conditions are of the evolution type 
--- we have to
specify initial data for them, otherwise their solutions are not
unique. Equations (18) for the perturbative contributions $\chi^{(n)}$ are
of different type --- in order to ensure uniqueness of their solution
it is sufficient to adopt the boundary conditions (17). The initial
data for $B(u^a)$ and $Y^i(u^a)$ follow from initial data for the
original field $\phi$. From such data for $\phi$ we know the
initial position and velocity of the core.  We assume that at the initial
instant $\tau_0$ the co-moving membrane
and the core have the same position and velocity. Hence,
\[
\mbox{initial data for the membrane = initial data for the core.}
\]
Using formula (5) one can show that then
\begin{equation}
B^{(n)}(\tau_0, \sigma^1, \sigma^2) = 0, \;\;\;
\partial_{\tau} B^{(n)}(\tau_0, \sigma^1, \sigma^2) = 0.
\end{equation}

In order to find the domain wall solution one should first solve the
collective dynamics, that is to compute evolution of the co-moving
membrane and of the $B^{(n)}$  fields. The profile $\chi$ of the domain wall
is found in the next step from formulas (15, 19).
In our perturbative
scheme the profile of the domain wall can not be chosen arbitrarily even
at the initial time --- it is fixed uniquely once the initial data
for the membrane and for the $B$ field are given.
Evolution of the core can be determined afterwards, from the
explicit expression for the scalar field $\phi$ \cite{8}.

\begin{center}
*   *   *
\end{center}

In paper \cite{8} the perturbative solution has been
constructed up to the fourth order.

In paper \cite{11}  second order perturbative solutions for
cylindrical and spherical domain walls have been compared with numerical
solutions of the Euler-Lagrange equation. The results of the 
comparison are quite encouraging.

In paper \cite{12} analogous approach has been applied to a
vortex in the Abelian Higgs model.

The perturbative approach to dynamics of the domain walls we
have sketched can be generalised to models involving several fields. Also
the requirement of relativistic invariance can be dropped out. \\

\vspace*{1cm}
{\bf Acknowledgment.} 
It is a pleasure to thank the organizers  for nice and
stimulating atmosphere  during the Workshop.


\begin{thebibliography}{99}

\bibitem{1} H.B.Nielsen and P.Olesen, Nucl.Phys. \underline{B61}, 45 (1973).

\bibitem{2} D.F\"{o}rster,  Nucl.Phys. \underline{B81}, 84 (1974).

\bibitem{3} R.Gregory,  Phys.Lett.\underline{B206}, 199 (1988).

\bibitem{4} S.M.Barr and D.Hochberg, Phys.Rev.\underline{D39}, 2308 (1989).

\bibitem{5} B.Carter and R.Gregory, Phys. Rev. \underline{D51}, 5839 (1995).

\bibitem{6} H.Arod\'{z} and P.W\c{e}grzyn, Phys.Lett.\underline{B291},
  251 (1992).
\bibitem{7} H. Arod\'{z},  Nucl. Phys. \underline{B450}, 174 (1995).
\bibitem{8} H. Arod\'{z}, Nucl. Phys. B, in press (hep-th/9703168).
\bibitem{9} B. Carter, in "Formation and Interactions of Topological
Defects", p.303. A.-Ch. Davis and R. Brandenberger (Eds.). Plenum Press,
New York and London, 1995.
\bibitem{10} N.G.van Kampen, Stochastic Processes in Physics and Chemistry.
North-Holland Publ.Comp., Amsterdam, 1987. Chapt.8,\S7.
\bibitem{11} J. Karkowski and Z. \'{S}wierczy\'{n}ski, Acta Phys. Pol. B
\underline{30}, 234 (1996).
\bibitem{12} H. Arod\'{z}, Nucl. Phys. \underline{B450}, 189 (1995).
\end{thebibliography}
\end{document}